\pdfoutput=1
\documentclass[aps,10pt,pre,superscriptaddress,byrevtex,twocolumn,bibnotes,footinbib,longbibliography]{revtex4-1}
\usepackage{microtype}
\usepackage{amsmath}
\usepackage{amssymb}
\usepackage{graphicx}
\usepackage{bm}
\usepackage{xcolor}
\usepackage{color}
\definecolor{myblue}{rgb}{0.153,0.322,0.706}
\usepackage[colorlinks,linkcolor=myblue,urlcolor=myblue,citecolor=myblue,breaklinks=true]{hyperref}

\newcommand{\be}{\begin{equation}}
\newcommand{\ee}{\end{equation}}
\newcommand{\ra}{\rightarrow}
\newcommand{\E}{\mathbb{E}}
\newcommand{\reals}{\mathbb{R}}
\newcommand{\p}{\partial}

\newcommand{\cA}{\mathcal{A}}
\newcommand{\cL}{\mathcal{L}}

\newcommand{\bsn}{\bm{\nabla}}
\newcommand{\bx}{\bm{x}}

\newcommand{\bX}{\bm{X}}
\newcommand{\bW}{\bm{W}}
\newcommand{\bF}{\bm{F}}
\newcommand{\bJ}{\bm{J}}

\newcommand{\hsigma}{\hat\sigma}

\newcommand{\bs}{\bm}
\newcommand{\sech}{\textrm{sech}}
\newcommand{\arctanh}{\textrm{arctanh}}
\newcommand{\transp}{\mathsf{T}}
\newcommand{\idmat}{\mathbb{I}}
\DeclareMathOperator{\Tr}{Tr}

\begin{document}

\title{Large deviations of the stochastic area for linear diffusions}

\author{Johan du Buisson}
\email{johan.dubuisson@gmail.com}
\affiliation{\mbox{Institute of Theoretical Physics, Department of Physics, Stellenbosch University, Stellenbosch 7600, South Africa}}

\author{Thamu D. P. Mnyulwa}
\affiliation{Department of Mathematical Sciences, Stellenbosch University, Stellenbosch 7600, South Africa}

\author{Hugo Touchette}
\email{htouchette@sun.ac.za}
\affiliation{Department of Mathematical Sciences, Stellenbosch University, Stellenbosch 7600, South Africa}

\date{\today}

\begin{abstract}
The area enclosed by the two-dimensional Brownian motion in the plane was studied by L\'evy, who found the characteristic function and probability density of this random variable. For other planar processes, in particular ergodic diffusions described by linear stochastic differential equations (SDEs), only the expected value of the stochastic area is known. Here, we calculate the generating function of the stochastic area for linear SDEs, which can be related to the integral of the angular momentum, and extract from the result the large deviation functions characterising the dominant part of its probability density in the long-time limit, as well as the effective SDE describing how large deviations arise in that limit. In addition, we obtain the asymptotic mean of the stochastic area, which is known to be related to the probability current, and the asymptotic variance, which is important for determining from observed trajectories whether or not a diffusion is reversible. Examples of reversible and irreversible linear SDEs are studied to illustrate our results.
\end{abstract}

\maketitle

\section{Introduction}

\begin{figure}
\includegraphics[width=0.34\textwidth]{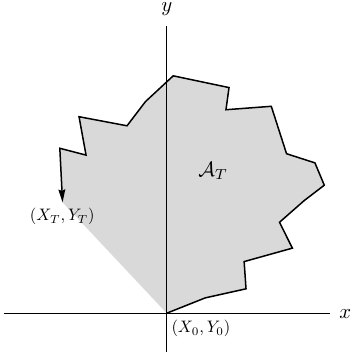}
\caption{Illustration of the stochastic area for planar Brownian motion started at the origin and finishing after a time $T$ at the point $(X_T,Y_T)$. $\cA_T$ is the area (in gray) enclosed by the path (here, not an actual path of Brownian motion) and the chord joining the final position to the origin. The arrow shows the direction of the path's evolution, leading to a positive stochastic area for the anticlockwise path shown.}
\label{fig1}
\end{figure}

Paul L\'evy studied in the 1940s and 50s \cite{levy1940b,levy1950,levy1951} the area enclosed by the paths of planar Brownian motion, expressed, in analogy with Green's theorem, by the line integral
\be
\cA_T= \frac{1}{2}\int_0^T X(t)dY(t)-Y(t)dX(t),
\label{eqareaito}
\ee
where $X(t)$ and $Y(t)$ are the coordinates of the Brownian motion; see Fig.~\ref{fig1}. This stochastic integral is interpreted as an It\^o integral and its sign is related in the usual way to the orientation of the path considered: positive for paths that rotate anticlockwise and negative for paths that rotate clockwise.

From the Fourier representation of Brownian motion, L\'evy calculated the characteristic function of $\cA_T$ \cite{levy1950}, obtaining
\be
\phi_T(k)=\E[e^{i k\cA_T}] = \sech\left(\frac{kT}{2}\right)
\label{eqlevycf}
\ee
from which we find the expected value $\phi'_T(0)=\E[\cA_T]=0$ and variance $\text{var}(\cA_T)=\phi''_T(0)=T^2/4$. Later, he realized that the characteristic function can be inverted to obtain the probability density of $\mathcal{A}_T$ \cite{levy1951}:
\be
p_T(a) =\frac{1}{T}\sech\left(\frac{\pi a}{T}\right),\quad a\in\reals.
\ee
This density is symmetric, which is consistent with planar Brownian motion being isotropic so that $\E[\cA_T]=0$, and also confirms the variance, which can be understood superficially by noting that, since the motion's radius $R$ typically grows as $\sqrt{T}$, we must have
\be
\text{var}(\cA_T)=\E[\cA_T^2]\sim R^4\sim T^2.
\ee

Recently, the stochastic area was re-examined from a more physical perspective in the context of linear stochastic differential equations (SDEs) that are ergodic \cite{ghanta2017,gonzalez2019,teitsworth2022}. For this class of Gaussian processes, an exact expression was found for the long-time expectation of $\mathcal{A}_T/T$, which turns out to be proportional to the antisymmetric component of the stationary probability current \cite{ghanta2017}. This is an interesting result, showing that $\cA_T$ can be used to determine from observed trajectories whether or not an SDE is reversible, that is, whether or not it satisfies the detailed balance condition with respect to its stationary distribution \cite{ghanta2017,gonzalez2019,teitsworth2022}. In this way, the stochastic area can be seen as an ``irreversibility metric'' \cite{gonzalez2019}, which is easier to measure or estimate as a scalar quantity than the stationary current, obtained in practice using force or velocity tracking experiments \cite{just2003,battle2016,perez-garcia2018,bruckner2020,frishman2020,gnesotto2018}, or the entropy production, which relies on detailed knowledge of the dynamics \cite{gnesotto2018,li2019,martinez2019,manikandan2020}.

In this paper, we continue these works by providing an exact form for the generating function of $\cA_T$ for linear ergodic SDEs, from which it is possible to obtain, in principle, its moments and its probability density at all times. By considering the long-time limit of the generating function, we also obtain explicit expressions for the large deviation functions characterizing the exponential form of the probability density of $\cA_T/T$ at large times. These functions have many applications in statistical physics, as they provide detailed information about the likelihood of fluctuations around stationary states, revealing in many systems the existence of fluctuation symmetries and dynamical phase transitions \cite{touchette2009,harris2013,jack2020}.

The large deviation functions of the stochastic area were also studied recently using path integral methods \cite{monthus2022}. Here, we derive these functions using a simpler method based on matrix Riccati equations \cite{buisson2022,buisson2022b} and use them to derive expressions for the asymptotic expectation of $\cA_T/T$, confirming the known result \cite{ghanta2017}, as well as the asymptotic variance, which is needed to validate the estimation of the expected area and to determine in experiments or simulations whether an SDE is irreversible. Two models are used to illustrate these results, namely, gradient SDEs, which are reversible, and a class of transverse SDEs that have a non-conservative drift, which makes them irreversible.

To complete our analysis, we also show how area fluctuations are created physically from rare trajectories that can be described by means of a modified SDE, called the effective or driven process \cite{jack2010b,chetrite2013,chetrite2014,chetrite2015}. The drift of this SDE is derived exactly using the Riccati formalism \cite{buisson2022,buisson2022b} and shows that non-zero (resp., zero) area fluctuations are created by trajectories that mimic an irreversible (resp., reversible) SDE, characterized by a modified current coupled in a non-trivial way to a modified density. These results are also illustrated with the two models above, and show in the end that the stochastic area can be used as a metric to characterize not only the irreversibility of the stationary state of a process, but also the irreversibility of its fluctuating trajectories. 

\section{Linear diffusions}

The systems that we consider are modelled as overdamped diffusions or Langevin-type systems that evolve according to the linear stochastic differential equation (SDE)
\be 
d\bX (t) = -M\bX(t) dt +\sigma d\bW(t),
\label{sde}
\ee
where $\bX(t) \in\reals^n$ is the state of the system at time $t$, $M$ is the $n\times n$ \emph{drift matrix} defining the linear deterministic force acting on the system, and $\bW(t)$ is an $m$-dimensional vector of independent Brownian motions acting as the noise source, which is multiplied by the $n \times m$ \emph{noise matrix} $\sigma$. To simplify the presentation, we consider the case $n=m=2$ of systems evolving in the $\reals^2$ plane, and discuss in the conclusion how our results generalize to systems evolving in $\reals^n$.

Linear SDEs are used to model many different systems, including noise-perturbed mechanical systems controlled by linear forces \cite{bechhoefer2021}, electrical circuits perturbed by Nyquist noise \cite{luchinsky1998}, and nonequilibrium systems driven by temperature or chemical baths \cite{gardiner1985}. One advantage of these SDEs is that the corresponding Fokker--Planck equation for the probability density $p(\bx,t)$ of $\bX(t)$ can be solved exactly for a fixed initial condition $\bX(0) = \bx_0$, yielding a Gaussian density for all times $t>0$; see, e.g., Sec.~3.7 of \cite{pavliotis2014}.

For our purposes, we require this density to have a long-time or stationary limit by assuming that the matrix $M$, which is not necessarily symmetric, is positive definite (i.e., has eigenvalues with positive real part) and, furthermore, that the diffusion matrix $D = \sigma \sigma^\transp$ is positive definite and invertible. Under these conditions, the SDE (\ref{sde}) is ergodic and has a unique (Gaussian) stationary density $p^*(\bx)$ \cite{pavliotis2014} given explicitly by
\be
p^*(\bx) = \frac{1}{2\pi \sqrt{\det C}} \exp\left(-\frac{1}{2}\langle \bx, C^{-1} \bx\rangle \right),
\label{pstat}
\ee
where $\langle \bs{a},\bs{b}\rangle$ denotes the standard vector inner product in $\reals^2$ and $C$ is the stationary covariance matrix satisfying the \emph{Lyapunov equation}
\be
D = MC + CM^\transp.
\label{lyapunov}
\ee
In addition, the SDE has a stationary probability current $\bJ^*(\bx)$ describing the long-time or average flow of probability in space. This current is defined generally as 
\be 
\bJ^*(\bx) = \bF(\bx) p^*(\bx) - \frac{D}{2} \bsn p^*(\bx), 
\label{curdef}
\ee
where $\bF(\bx)$ is the deterministic force or \emph{drift} entering in the SDE. In our case, $\bF(\bx) = -M\bx$, which yields, together with (\ref{pstat}), 
\be
\bJ^*(\bx) = H\bx p^*(\bx),
\label{statcurlin}
\ee
where 
\be
H = \frac{DC^{-1}}{2}  - M.
\label{hmat}
\ee

The knowledge of $p^*$ and $\bJ^*$ is important physically as it determines whether or not a process is reversible, that is, whether or not the process has the same statistics when its evolution is reversed in time. This reversibility property is defined mathematically via the detailed balance condition and is related to the current in the following way: if $\bJ^*(\bx)=\bs{0}$ for all $\bx$, then the SDE is reversible, describing in the long-time limit an equilibrium steady state, whereas if $\bJ^*(\bx)\neq \bs{0}$, then the SDE is irreversible and describes a nonequilibrium steady state violating the condition of detailed balance \footnote{This classification applies for overdamped diffusions. For underdamped diffusions, there can be a current in phase space even at equilibrium.}.

For linear SDEs, we can distinguish in general two sources of nonequilibrium behavior, namely, a non-symmetric drift matrix $M$, indicating the presence of a non-conservative force, and a diffusion matrix $D$ that is not proportional to the identity matrix $\idmat$, which arises when a system is in contact with multiple heat baths at different temperatures or when correlations exist between the noise sources \cite{zia2007,kwon2011,noh2014}. Though distinct physically, these two sources can cancel each other in some cases to create a reversible steady state if $M$ and $D$ are such that $H = 0$.

\section{Stochastic area}

Let $(\bX(t))_{t=0}^T$ be a path of the linear SDE \eqref{sde} in $\reals^2$ over the time interval $[0,T]$ with components $\bX(t)=(X(t),Y(t))$.  Our aim is to study the statistics of the stochastic area enclosed by this path (see Fig.~\ref{fig1}), which we divide by $T$ so as to consider the \emph{time-averaged area} or \emph{area per unit time}
\be
A_T= \frac{\cA_T}{T}=\frac{1}{2T}\int_0^T X(t)\circ dY(t) - Y(t)\circ dX(t),
\label{eqareastrat}
\ee
where $\circ$ denotes the Stratonovich or mid-point convention for the stochastic integral. Alternatively, we can write
\be
 A_T = \frac{1}{T} \int_0^T \Gamma \bX(t) \circ d\bX(t),
\label{stochobs}
\ee
where $\Gamma$ is the \emph{antisymmetric} matrix
\be
\Gamma = \frac{1}{2} \begin{pmatrix}0 && -1 \\ 1 && 0 \end{pmatrix}
\ee
and $\circ$ now denotes, with a slight abuse of notation, the scalar product interpreted in the Stratonovich convention.

The stochastic area can be interpreted physically as an integrated angular momentum or torque \cite{gonzalez2019} and is an example in stochastic thermodynamics of \emph{current-type observables}, which can be related in many cases to physical quantities, such as the mechanical work or the entropy production \cite{seifert2012}. The connection with the current comes from the fact that the stationary expectation of $A_T$ is given by
\be
a^* = \lim_{T\ra\infty} \E[A_T]=\int_{\reals^2} \langle \Gamma \bx, \bJ^*(\bx)\rangle\, p^*(\bx)\, d\bx
\label{statvaldef}
\ee
and so involves not only the stationary density $p^*$, given in (\ref{pstat}), but also the stationary current $\bJ^*$, shown in (\ref{statcurlin}). Using these expressions, the integral above reduces in form to the second moment of a Gaussian density, yielding the trace formula \cite{buisson2022,buisson2022b}
\be
a^* = \Tr\left(\Gamma^{\transp}HC\right) .
\label{statval}
\ee
From the Lyapunov equation \eqref{lyapunov}, it is easy to show that the matrix $HC=D/2-MC$ is antisymmetric, so the trace can be calculated exactly with $\Gamma$ to obtain
\be
a^*= \left(MC-\frac{D}{2}\right)_{1,2},
\label{eqareaghanta}
\ee 
which recovers the result obtained before \cite{ghanta2017}. Moreover, from the expression of the stationary current, we can write
\be
\bJ^*(\bx)= 2a^*\Gamma C^{-1}\bx p^*(\bx)
\ee
in two dimensions, showing in this case that a linear diffusion is reversible if and only if $a^*=0$. 

The value $a^*$ is not only the stationary expectation of $A_T$, but also the typical value of this random variable most likely to be observed in experiments or simulations when the integration or ``observation'' time $T$ is large. This follows from the ergodic theorem generalized to stochastic integrals, which states for ergodic diffusions that $A_T$ converges in probability to $a^*$ as $T\ra\infty$ \cite{chetrite2014}. We study in the next section this convergence more precisely by determining the distribution of $A_T$ around $a^*$. 

To complete this section, it is useful to note that $A_T$ is also an example in mathematics of integrals of differential forms along stochastic paths, referred to as \emph{stochastic currents}, which have been studied since the 1970s \cite{yor1977,ikeda1979,ochi1985,manabe1992,kuwada2003,kuwada2006}. In this context, it is common to define stochastic integrals using the Stratonovich convention, as done in \eqref{eqareastrat}. However, in the particular case of the stochastic area, it is interesting to note that the stochastic convention or calculus used for calculating the integral is irrelevant as \emph{all conventions give the same result}, essentially because this quantity is a linear antisymmetric differential form; see Sec.~4.2 of \cite{mnyulwa2023} for the proof. Therefore, all conventions agree on what the area enclosed by a path is, as expected geometrically and physically, which explains why the result for $a^*$ in \eqref{eqareaghanta} is the same whether it is derived using  the It\^o convention \cite{ghanta2017} or, as done here, using the Stratonovich convention.

\section{Area statistics}

We derive in this section our main results for the statistics of $A_T$ for linear diffusions using a general formalism developed recently for linear diffusions \cite{buisson2022,buisson2022b}. We focus on the long-time limit of the probability density $p(A_T=a)$, which shows a large deviation form that explains the convergence of $A_T$ towards the stationary expectation $a^*$, with fluctuations around this value that are exponentially unlikely with $T$. Using the same formalism, we describe the paths leading to these fluctuations in terms of a modified diffusion, which changes in general the reversible or irreversible nature of the diffusion considered.

\subsection{Generating function}

To our knowledge, the probability density of the mean area cannot be obtained in closed form directly for general linear diffusions, so we consider instead its generating function, defined as 
\be
G_k(\bx,t) = \E\left[e^{k t A_t}|\bX(0) = \bx \right].
\ee
This can be viewed as the Laplace transform of the conditional probability density $p(A_t = a|\bX(0) = \bx)$, up to a scaling factor involving the time $t$. Provided that the generating function exists in a neighbourhood of $k = 0$, we can extract the moments of $A_T$ as 
\be
\E\left[(tA_t)^n|\bX(0) = \bx\right] = \left.\frac{d^n G_k(\bx,t)}{dk^n}\right|_{k = 0}.
\label{moments}
\ee

To find the generating function, we use the fact that its time evolution satisfies the linear partial differential equation
\be
\p_t G_k(\bx,t) =\cL_k G_k(\bx,t), \quad G_k(\bx,0) = 1,
\label{FKeq}
\ee
known as the \emph{Feynman--Kac} (FK) equation \cite{stroock1979}. In this equation, $\cL_k$ is a differential operator, known as the \emph{tilted generator}, whose expression depends on the generator of the process $\bX(t)$ and the observable under consideration (see Eq.~(29) in \cite{touchette2017}). In our case, we have
\begin{align}
\cL_k &= -k\langle M\bx,\Gamma\bx\rangle + \langle(-M + kD\Gamma)\bx,\bsn\rangle  \nonumber\\ 
   & \qquad + \frac{1}{2} \langle\bsn, D \bsn\rangle + \frac{k^2}{2} \langle\bx,\Gamma^{\transp}D\Gamma\bx\rangle.
\end{align}

Recently, an exact expression for the solution of the FK equation was found for general linear current-type observables of linear ergodic diffusions \cite{buisson2022,buisson2022b}. When applied to the stochastic area, this expression gives
\be
G_k(\bx,t) = \exp\left(\langle \bx, B_k(t) \bx \rangle + \int_0^t \Tr[D B_k(s)] ds \right),
\label{genfun}
\ee
where $B_k(t)$ is a $2 \times 2$ symmetric matrix that satisfies the following differential Riccati equation:
\begin{align}
\frac{dB_k(t)}{dt} &= \frac{k^2}{2} \Gamma^{\transp}D\Gamma + 2B_k(t) DB_k(t) - \frac{k}{2}(M^{\transp}\Gamma-\Gamma M) \nonumber \\ 
& \quad+ (-M + kD\Gamma)^{\transp}B_k(t) + B_k(t)(-M + kD\Gamma),
\label{riccati}
\end{align}
with initial condition $B_k(0) = 0$. We refer to  Sec.~III.C.1 in \cite{buisson2022b} for the derivation of this result.

The matrix Riccati equation can be solved exactly for some choices of $M$ and $D$, as we show in the next section, to obtain an explicit expression for $G_k(\bx,t)$, which can then be used to obtain the moments of $A_T$ by differentiation or its density by performing an inverse Laplace transform (either analytically or numerically). In cases where it cannot be solved exactly, one can also resort to numerical techniques developed for Riccati equations to obtain numerical estimates of $G_k(\bx,t)$, which can be further differentiated or inverted numerically. 

Our experience of this equation suggests, however, that little insight is gained by computing $B_k(t)$ accurately as a function of time, especially if one is interested in the statistics of $A_T$ for large times. This is because this matrix quickly converges to a fixed point in most cases of interest, which implies that $G_k(\bx,t)$ scales exponentially with $t$ as $t\ra\infty$ and, in turn, that $p(A_T=a)$ scales exponentially with $T$ as $T\ra\infty$. These results are studied next using large deviation theory.

\subsection{Large deviations}

To study the long-time form of the generating function, we assume that the differential Riccati equation (\ref{riccati}) has a stationary solution $B_k^*$, satisfying the algebraic Riccati equation 
\begin{align} 
0 & =\frac{k^2}{2} \Gamma^{\transp}D\Gamma - \frac{k}{2}(M^{\transp}\Gamma-\Gamma M) + 2B_k^* DB_k^* \nonumber \\ 
    &\qquad+ (-M + kD\Gamma)^{\transp}B_k^* +B_k^*(-M + kD\Gamma)
\label{Bkcur}
\end{align} 
with the requirement $B_0^* = 0$. In this case, we see from \eqref{genfun} that $G_k(\bx,t)$ scales asymptotically for large times as 
\be
G_k(\bx,t) \approx r_k(\bx) e^{t\lambda(k)},
\label{longtimegen}
\ee
where 
\be
\lambda(k) = \Tr\left(DB_k^*\right)
\label{scgfstoch}
\ee
and 
\be
r_k(\bx) = e^{\langle \bx, B_k^*\bx\rangle}.
\label{rkstoch}
\ee

The function $\lambda(k)$ is known in large deviation theory as the \emph{scaled cumulant generating function} (SCGF) and determines the long-time scaling of $p(A_T=a)$ via the G\"{a}rtner-Ellis theorem which states that, if $\lambda(k)$ exists and is differentiable, then
\be 
p(A_T = a) \approx e^{-T I (a)},
\label{LDP}
\ee
with corrections that are sub-exponential in time. Moreover, the \emph{rate function} $I(a)$ that controls the exponential decay is given by the Legendre transform of the SCGF: 
\be
I(a) = k_a a - \lambda(k_a),
\label{leg}
\ee
$k_a$ being the unique solution of 
\be
\lambda'(k) = a. 
\label{legdiff}
\ee

For ergodic linear diffusions, and ergodic Markov processes in general, the rate function is convex, positive, and has a unique minimum at the stationary expectation $a^*$, which implies that the density $p(A_T =a)$ concentrates exponentially on $a^*$, so that this value is the typical value of $A_T$, as mentioned before. In general, we have $a^*=\lambda'(0)$, which recovers with \eqref{scgfstoch} and \eqref{Bkcur} the result shown in \eqref{statval}. Similarly, the asymptotic variance of $A_T$ is given by $\lambda''(0)$, so that
\be
\text{var}[A_T]\sim \frac{\lambda''(0)}{T}.
\label{asympvar}
\ee
From \eqref{scgfstoch} and \eqref{Bkcur}, we obtain explicitly
\begin{align} 
\lambda''(0) &= \Tr\left[C\Gamma^{\mathsf{T}}D\Gamma + 2 C(\Gamma M - M^{\mathsf{T}}\Gamma) CB_0^{*\prime}\right]\nonumber \\
   	&  \qquad + \Tr\left[2 C \left(\Gamma^{\mathsf{T}}D B_0^{*\prime} + B_0^{*\prime}D\Gamma\right)\right],
\label{curvar}
\end{align}
where $B_0^{*\prime}$ is the derivative of $B_k^*$ with respect to $k$ evaluated at $k = 0$, which satisfies the Lyapunov equation 
\be 
B_0^{*\prime}M + M^{\mathsf{T}}B_0^{*\prime} = \frac{1}{2}\left(\Gamma M - M^{\mathsf{T}}\Gamma \right).
\label{bigricc1}
\ee
In principle, higher order cumulants can be calculated in a similar manner.

\subsection{Effective process}

The rate function determines at an exponential scale the likelihood of fluctuations of $A_T$ around its typical value $a^*$. To understand how these fluctuations arise dynamically, we can imagine performing an experiment where we single out trajectories that have the same area value, say $A_T=a$, and analyze this subset of trajectories to determine whether they can be described as a Markov process, in our case with an SDE. Mathematically, this experiment corresponds to conditioning the process $\bX(t)$ on the event $A_T=a$, and is known from recent works \cite{jack2010b,chetrite2013,chetrite2014,chetrite2015} to lead in the long-time limit to a modified Markov process, referred to as the \emph{effective}, \emph{driven} or \emph{fluctuation process}, which describes the dynamics of the conditioned set of trajectories. To be more precise, those trajectories, which are rare in the original process, correspond in the long-time limit to the trajectories of an effective process that realize the value $A_T=a$ in a typical way (see \cite{chetrite2013,chetrite2014,chetrite2015} for more details).

In our case, the effective process takes the form of a modified SDE, expressed as
\be
d\bX_k(t) = \bF_k(\bX_k(t))dt +\sigma d\bW(t),
\label{effsde}
\ee
which has the same noise matrix as the original SDE in \eqref{sde}, but now involves a modified drift $\bF_k(\bx)$, called the \emph{effective drift}, given by
\be 
\bF_k(\bx) = -M_k \bx,
\label{stoceffdrift}
\ee
where
\be
M_k = M - kD\Gamma -2DB_k^*
\label{effmk}
\ee
is a modification of the drift matrix $M$. Here, the parameter $k$ is set according to the Legendre relation in \eqref{legdiff}, which plays a role similar to the temperature-energy relation in equilibrium statistical mechanics \cite{touchette2009}, in that $k$ can be seen as an inverse temperature that is tuned to achieve a fixed value $A_T=a$ of the stochastic area. In particular, for $k=0$ we have $\bF_0(\bx)=-M\bx$, the drift of the original SDE in \eqref{sde}, since $\lambda'(0)=a^*$.

We refer to our recent work on the large deviations of linear diffusions \cite{buisson2022,buisson2022b} for the derivation of these results. For our purposes, the main point to note is that the effective SDE \eqref{effsde} preserves the linearity of the original SDE \eqref{sde} because the stochastic area is a linear current-type observable. As a result, it is characterized in the long-time limit by a Gaussian stationary probability density, given by 
\be 
p_k^*(\bx) = \frac{1}{2\pi \sqrt{\det C_k}} \exp\left(-\frac{1}{2} \langle\bx, C_k^{-1}\bx\rangle \right),
\label{statdeneff}
\ee
where $C_k$ is now the unique positive-definite solution to the Lyapunov equation 
\be
D=M_kC_k + C_k M_k^{\transp}.
\label{lyapunov2}
\ee
Moreover, it has a modified stationary current $\bJ_k^*$ given, similarly to (\ref{statcurlin}), by
\be 
\bJ_k^*(\bx) = H_k\bx p_k^*(\bx),
\label{statcureff}
\ee
with $H_k$ given from (\ref{hmat}) with the substitutions $C \rightarrow C_k$ and $M \rightarrow M_k$ as
\be
H_k = \frac{D C_k^{-1}}{2} - M_k.
\label{hkmat}
\ee
These results are again expressed in terms of $k$. To relate them to a given area value $A_T=a$, we need to fix $k$ so that \eqref{legdiff} is satisfied. Alternatively, we can note that the effective process must realize a typical value $a^*_k$ of the area for a given $k$, obtained from \eqref{statvaldef} by replacing $p^*$ with $p^*_k$  and $\bJ^*$ with $\bJ_k^*$. This yields
\be
a^*_k = \Tr\left(\Gamma^{\transp}H_kC_k\right),
\ee
which is consistent with $\lambda'(k)=a^*_k$ and the fact that the trajectories leading to $A_T=a$ are realized as the typical trajectories of the effective SDE.

In the next section, we study the manner in which the stationary density and current of the modified SDE differ from those of the original SDE in order to understand the manner in which the latter process must behave at the level of its fluctuating trajectory to manifest a particular area fluctuation. Given the nature of this observable, it should be clear already that non-zero area fluctuations will be associated with a non-zero effective current, so the reversible nature of a process can be modified at the level of fluctuations. 

\section{Applications}

We illustrate our results in this section for two simple linear diffusions that are representative of reversible and irreversible diffusions. We obtain for both the exact generating function and extract from it the mean and variance, as well as the SCGF and rate function describing the area fluctuations in the long-time limit. Finally, we analyze the form of the effective drift to identify the physical mechanisms that create these fluctuations in the presence of conservative and nonconservative forces.

\subsection{Gradient diffusion}

The first model that we consider is a diffusion $\bX(t)$ in $\reals^2$ satisfying the SDE (\ref{sde}) with the drift matrix
\be
M = \begin{pmatrix}\gamma && 0 \\ 0 && \gamma \end{pmatrix} = \gamma \idmat
\ee
with $\gamma >0$ and noise matrix $\sigma = \epsilon \idmat$ with $\epsilon>0$, so that $D = \epsilon^2 \idmat$. This process describes, as is well known, the dynamics of an overdamped Brownian particle in a quadratic potential with friction or ``stiffness'' $\gamma$ \cite{risken1996}, characterized by the stationary density
\be
p^*(\bx) =\frac{\gamma}{\pi\epsilon^2}\exp\left(-\frac{\gamma}{\epsilon^2}\|\bx\|^2 \right),
\label{gradstatden}
\ee
where $\|\bx\|=\langle\bx, \bx\rangle$, and a stationary current that vanishes everywhere, since the drift is conservative, meaning that it is the gradient of a potential. Therefore, the process is reversible, as confirmed by the fact that $a^*=0$.

To determine the distribution of $A_T$ around this typical value, we consider its generating function, as given by \eqref{genfun}, noting that, for the system considered, the matrix $B_k(t)$ is proportional to the identity matrix for all times $t$, as shown in Sec.~5.2.1 of \cite{buisson2022}, so we write 
\be 
B_k(t) = b_k(t) \idmat
\label{diaggrad}
\ee
where $b_k(t)$ is a scalar function. In this case, the differential Riccati equation \eqref{riccati} becomes an ordinary differential equation for $b_k(t)$, taking the form
\be 
\frac{db_k(t)}{dt} = \frac{\epsilon^2 k^2}{8} + 2\epsilon^2 b_k^2(t) - 2\gamma b_k(t), \quad b_k(0) = 0,
\label{diffeq2}
\ee
having substituted the relevant expressions for $M$ and $D$ into (\ref{riccati}). This equation can be solved explicitly to obtain
\be 
b_k(t) = \frac{\gamma}{2 \epsilon^2} \left(1 - \frac{1 + \frac{\gamma_k}{\gamma} \tanh\left(t\gamma_k \right)}{1 + \frac{\gamma}{\gamma_k} \tanh\left(t \gamma_k \right)} \right),
\label{bkfull}
\ee
where 
\be
\gamma_k = \sqrt{\gamma^2 - \frac{k^2 \epsilon^4}{4}}.
\label{eqeffgamma1}
\ee
Moreover, the integral of the trace in \eqref{genfun} can be calculated explicitly, yielding the following expression for the generating function:
\begin{align} 
G_k(\bx,t) &=e^{b_k(t)\|\bx\|^2} \nonumber \\ 
&\quad \times\frac{ \sqrt{8}\gamma_k \exp\left(\gamma t - \arctanh\left[\frac{\gamma}{\gamma_k} \tanh(t\gamma_k) \right] \right)}{\sqrt{8\gamma^2 - k^2\epsilon^4[1 + \cosh(2t \gamma_k)]}},
\label{exactres}
\end{align}
valid for $k \in (-2\gamma/\epsilon^2, 2\gamma/\epsilon^2)$. It can be checked that this recovers L\'evy's result in \eqref{eqlevycf} for Brownian motion by setting $\gamma=0$ and $\epsilon=1$, and by replacing $k$ by $ik$ to obtain the characteristic function rather than the generating function.

It is unclear whether the expression in \eqref{exactres} can be inverted by Laplace transform to obtain an analytic expression for the density $p(A_t = a )$ starting from the initial condition $\bX(0)=\bx$. We have tried to invert it numerically, but the result yields little information graphically beyond the expectation and variance which can be computed analytically from \eqref{moments}, obtaining
\be
\E[A_T] = 0
\ee
for all times $T$, starting from $\bX(0) =\bx$, and
\begin{align}
\text{var}[A_T] &=\frac{e^{-\gamma T}}{4\gamma^2 T^2} \bigg[ \gamma\epsilon^4  T\cosh(\gamma T)\nonumber \\
&\qquad +\epsilon^2\left[\|\bx\|^2 \gamma + \epsilon^2 (\gamma T - 1) \right]\sinh(\gamma T)\bigg]
\label{gradvar}
\end{align}
starting from the same initial condition. The first result is expected intuitively given that the gradient dynamics has no inherent rotational bias. As for the variance, it can be checked that \eqref{gradvar} decays to leading order like $1/T$, confirming that the density $p(A_T=a)$ concentrates on $a^*=0$ as $T$ increases.

This concentration is determined again by the large deviation approximation \eqref{LDP}. To find the associated rate function, we note that the stationary solution $B_k^*$ to the Riccati equation (\ref{riccati}) can be obtained either by taking the limit $t \rightarrow \infty$ in (\ref{bkfull}) or by solving the algebraic Riccati equation (\ref{Bkcur}), given here for the coefficient $b_k^*$ as 
\be
\frac{\epsilon^2 k^2}{8} + 2 \epsilon^2 {b_k^*}^2 - 2\gamma b_k^* = 0,
\ee
and by choosing the solution with $b_0^* =0$. The result of either procedure is 
\be 
b_k^* =\frac{\gamma-\gamma_k}{2\epsilon^2},
\label{bkstat}
\ee
so that the SCGF is found from (\ref{scgfstoch}) to be 
\be 
\lambda(k) = \gamma - \gamma_k
\label{scgfstocgrad}
\ee
for $k$ in the interval shown after \eqref{exactres}. Taking the Legendre transform of this expression, we then find
\be
I(a)=\frac{\gamma \epsilon^2}{\sqrt{4 a^2+\epsilon ^4}}+\frac{4 a^2 \gamma }{\sqrt{4 a^2\epsilon ^4+\epsilon ^8}}-\gamma
\ee
for the rate function, defined for all $a\in\reals$.

\begin{figure}[t]
\centering
\includegraphics{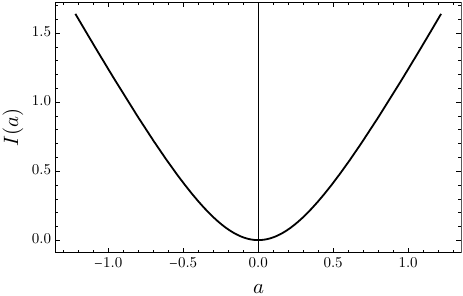}
\caption{Rate function for the mean stochastic area of the gradient diffusion for the parameter values $\gamma = 1$ and $\epsilon=1$.}
\label{fig:gradscgfrate}
\end{figure}

The plot of this function in Fig.~\ref{fig:gradscgfrate} shows that there are two regimes of area fluctuations. Close to the typical value $a^*=0$, $I(a)$ is locally quadratic, which implies, on the one hand, that the \emph{small} fluctuations or deviations of $A_T$ close to this value are Gaussian-distributed with a variance that decays according to \eqref{asympvar} with an asymptotic variance equal to
\be
\lambda''(0) = \frac{\epsilon^4}{4\gamma}.
\ee
This value is important, as we discuss in the next section, for determining the error bars associated with the estimation of the expected area. On the other hand, far from $a^*$, $I(a)$ scales according to \be
I(a) \sim \frac{2\gamma}{\epsilon^2}|a|
\ee
so the \emph{large} deviations of the stochastic area are exponentially-distributed. This can also be inferred from the fact that $\lambda(k)$ is defined for a limited range of values $k\in (k_-,k_+)$ around the origin. In this case, the left tail of $I(a)$ has an asymptotic slope equal to $k_-$ while its right tail has an asymptotic slope equal to $k_+$; see Example 3.3 in \cite{touchette2009} for more details. 

\begin{figure}[t]
\centering
\includegraphics{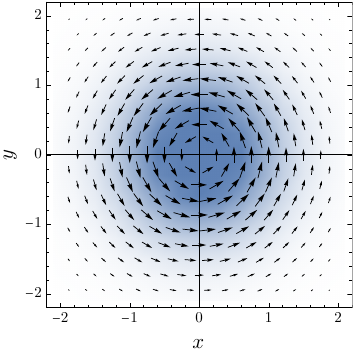}
\caption{Vector plot of the stationary current $\bJ_k^*(x,y) $ of the effective process associated with the stochastic area for the gradient diffusion. The density plot underneath shows the modified stationary density $p_k^*(x,y)$. Parameters: $\gamma =1, \epsilon=1$, and $k = 1$, corresponding to the positive fluctuation $a=\lambda'(1)=1/(2\sqrt{3})$. For negative fluctuations $a<0$, the plot is very similar; all that changes is the orientation of the current, which switches from anticlockwise for $k>0$ to clockwise for $k<0$. For $k=0$, $\bJ_{0}^*=\bJ^*=\mathbf{0}$.}
\label{fig:gradstocarea}
\end{figure}

To understand the mechanisms at play behind these two types of fluctuations, we consider the effective drift matrix $M_k$ characterizing the linear effective process with drift \eqref{stoceffdrift}. From the expressions of $\gamma$, $D$ and $B_k$, we find from \eqref{effmk}
\be  
M_k = \begin{pmatrix}\gamma_k && k \epsilon^2/2 \\ -k\epsilon^2/2 &&\gamma_k \end{pmatrix},
\label{mkgrad}
\ee
which is positive definite for $k$ in the range shown after (\ref{exactres}). This drift matrix defines a so-called transverse system, characterized by an effective stiffness $\gamma_k$ in the diagonal, which determines with \eqref{statdeneff} and \eqref{lyapunov2} the stationary density
\be
p_k^*(\bx) = \frac{\gamma_k}{\pi\epsilon^2}\exp\left(-\frac{\gamma_k}{\epsilon^2}\|\bx\|^2\right),
\ee
and a transverse, off-diagonal term $k\epsilon^2/2$ that creates a rotation in the dynamics, leading from \eqref{statcureff} and \eqref{hkmat} to a non-zero stationary current
\be
\bJ_k^*(\bx) = \frac{k\epsilon^2}{2}\begin{pmatrix}-y \\ x \end{pmatrix}p_k^*(\bx)
\label{effcurgrad}
\ee
when $k\neq 0$. This current rotates in an anticlockwise circular direction for $k > 0$ and a clockwise circular direction for $k < 0$, as shown in Fig.~\ref{fig:gradstocarea}.

Intuitively, for a positive fluctuation $a > 0$ to occur, the system can perform more anticlockwise loops than clockwise loops around the origin or, once a bias for anticlockwise loops already exists, perform anticlockwise loops that venture far away from the origin so as to accumulate more area. From the result above, we see that the gradient diffusion exploits both of these strategies when fluctuating: the non-conservative rotational force in the effective drift increases in magnitude with $k$ to create a current that favors rotating trajectories, while the effective friction $\gamma_k$ decreases with $k$, letting the process spend more time further away from the origin, thereby creating larger area loops around the origin. This is illustrated in Fig.~\ref{fig:gradstocarea} for the parameter values $\gamma=1$, $\epsilon=1$ and $k=1$ associated with a positive area fluctuation and thus to an anticlockwise stationary current $\bJ_k^*(\bx)$.

The same picture holds for negative area fluctuations, associated with $k<0$. In this case, the direction of the current is reversed, forcing clockwise rotating trajectories, which are confined, however, in the same way around the origin by the friction $\gamma_k$, since it is even in $k$.

Interestingly, as $k\ra \pm2\gamma/\epsilon^2$, the two boundary values for which the SCGF is defined, we see from \eqref{eqeffgamma1} that $\gamma_k\ra 0$, so the confinement gets weaker for large area fluctuations (either positive or negative), while the rotation parameter $k\epsilon^2/2$ saturates to $\pm\gamma$. This means that the effective process favors trajectories that creates large loops over trajectories that rotates faster. In other words, if a large, say positive, area fluctuation is observed, then it is more likely that this area was created by a trajectory that ventured far away from the origin (anticlockwise around the origin) as a opposed to a trajectory that rotates very fast (anticlockwise) close to the origin. This density effect gets stronger as $|k|$ or $|a|$ increases, which explains the crossover from Gaussian to exponential fluctuations \cite{tsobgni2016,nardini2018}.

\subsection{Transverse diffusion}

For our second application, we consider an irreversible system that sustains a non-zero stationary current, such that $a^*\neq 0$. To this end, we take the drift and diffusion matrices to be
\be
M = \begin{pmatrix}\gamma && \xi \\ -\xi && \gamma \end{pmatrix}
\ee
and $D = \epsilon^2 \idmat$, respectively. This system is again a transverse diffusion, which serves in physics as a minimal model of irreversible processes \cite{zia2007,kwon2011,noh2014,volpe2006}. As before, the antisymmetric part of the drift involving the parameter $\xi$ creates a stationary current given by
\be 
\bJ_k^*(\bx) = \xi \begin{pmatrix}-y \\ x \end{pmatrix}p_k^*(\bx),
\label{Jtrans}
\ee
with the stationary density 
\be 
p^*(\bx) =\frac{\gamma}{\pi\epsilon^2}\exp\left(-\frac{\gamma}{\epsilon^2}\|\bx\|^2 \right),
\label{ptrans}
\ee
which is the same as that obtained in (\ref{gradstatden}) for the gradient diffusion. The current $\bJ^*$ is anticlockwise for $\xi>0$ and clockwise for $\xi<0$, and leads with (\ref{statval}) to the following typical value of the stochastic area:
\be
a^* = \frac{\epsilon^2 \xi}{2\gamma},
\label{asympmeantrans}
\ee
which is positive for $\xi > 0$ and negative for $\xi <0$.

The generating function of the stochastic area for this diffusion can also be found exactly, as $B_k(t)$ is still proportional to the identity, and leads in fact to the same result as in \eqref{exactres}, but with
\be
\gamma_k = \sqrt{\gamma^2 - \frac{k\epsilon^2(k\epsilon^2 + 4\xi)}{4}}
\label{lambdanew}
\ee
for $k\in (k_-,k_+)$ where
\be 
k_\pm =\frac{-2\xi \pm 2\sqrt{\gamma^2+\xi^2}}{\epsilon^2}.
\label{range}
\ee
The change of $\gamma_k$ affects the moments of $A_T$, including its expectation, which is now 
\be 
\E[A_T] = \frac{\xi\left(\gamma \epsilon^2 T + [\gamma\|\bx\|^2 + \epsilon^2(\gamma T - 1)\tanh(\gamma T) ] \right)}{2\gamma^2T[1 + \tanh(\gamma T)]},
\label{timedepmean2}
\ee
starting from $\bX(0)=\bx$. Moreover, it can be checked from \eqref{lambdanew} and \eqref{exactres} that $G_k(\bx,t)$ is invariant under the replacement $k \rightarrow -k - 4\xi/\epsilon^2$ for all times $t$. This is a well-known symmetry of the generating function, referred to as the \emph{fluctuation relation} \cite{harris2007}, which implies at the level of the distribution of $A_T$ that
\be
\frac{p(A_T = a)}{p(A_T= -a)} = e^{4\xi a T/\epsilon^2},
\label{FR}
\ee
so that fluctuations with a particular sign are exponentially suppressed with $T$ relative to fluctuations with the opposite sign. Such a symmetry is normally associated with the entropy production \cite{seifert2012} and arises here because the stochastic area can be related for linear diffusions in two dimensions to the nonequilibrium work, which is itself related to the entropy production \cite{buisson2022b}. 

The fluctuation relation can also be inferred from the large deviation functions. Since the form of $G_k(\bx,t)$ is unchanged compared to the first model, the SCGF is still given by \eqref{scgfstocgrad} but for $k$ in the range \eqref{range} and $\gamma_k$ in \eqref{lambdanew}. From this, the rate function is obtained by Legendre transform as
\begin{align}
I(a) &= 
\sqrt{\frac{\epsilon ^4 \left(\gamma ^2+\xi ^2\right)}{4 a^2+\epsilon ^4}}-\frac{2 a \xi }{\epsilon^2}-\gamma\nonumber\\
&\quad+\frac{4 |a|^3 \left(\gamma ^2+\xi ^2\right) }{\sqrt{a^2 \epsilon ^4 \left(4a^2+\epsilon ^4\right) \left(\gamma ^2+\xi ^2\right)}},
\end{align}
which is such that
\be
I(-a) = I(a) +\frac{4\xi a}{\epsilon^2},
\ee
consistently with \eqref{FR}.

\begin{figure}[t]
\centering
\includegraphics{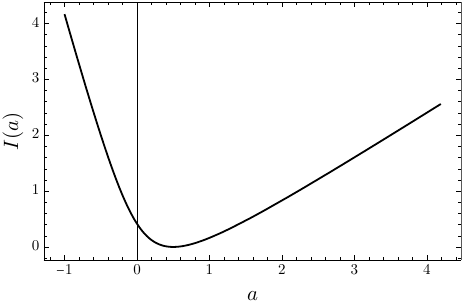}
\caption{Rate function for the stochastic area of the transverse diffusion for the parameter values $\gamma = 1,\xi=1$, and $\epsilon=1$, for which $a^*=0.5$.}
\label{fig:fig3}
\end{figure}

\begin{figure*}
\centering
\includegraphics[width=0.95\textwidth]{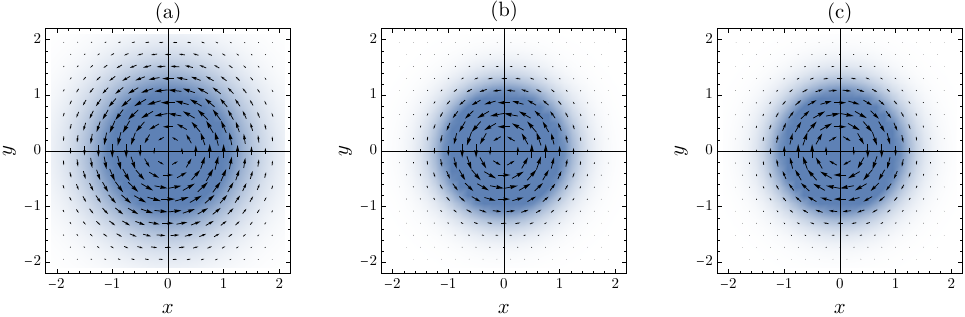}
\caption{Vector plot of the stationary current $\bJ_k^*(x,y)$ of the effective process associated with the stochastic area for the transverse diffusion for the three fluctuation regimes: (a) $a>a^*$, (b) $0<a<a^*$, (c) $a<0$. The density plots underneath show the modified stationary density $p_k^*(x,y)$. Parameters: $\gamma =1$, $\xi=1$, and $\epsilon=1$.}
\label{fig:fig4}
\end{figure*}

The plot of $I(a)$ is shown in Fig.~\ref{fig:fig3} for $\gamma=\xi=\epsilon=1$. Compared with the gradient diffusion, we now see that $I(a)$ is asymmetric around the typical value $a^*$, indicating that the system is more likely to produce fluctuations that go in the direction of the current. Close to $a^*$, we still have a Gaussian regime of fluctuations characterized by the asymptotic variance
\be
\lambda''(0) =\frac{\epsilon^4(\gamma^2+\xi^2)}{4\gamma^3}.
\ee
Moreover, the tails of $I(a)$ are also asymptotically linear, implying that the large deviations of $A_T$ are still exponentially distributed, though in an asymmetric way, so that we have
\be
I(a)~\sim \frac{-2\xi -2\sqrt{\gamma^2+\xi^2}}{\epsilon^2}a
\ee
as $a \rightarrow -\infty$ and 
\be
I(a) \sim \frac{-2\xi +2\sqrt{\gamma^2+\xi^2}}{\epsilon^2}a
\ee
as $a \rightarrow \infty$.

From these results, we expect the effective diffusion to behave, as in the case of the gradient diffusion, as a modified transverse diffusion that adjust the stiffness and rotation parameters so as to produce specific area fluctuations. This is confirmed by calculating $M_k$ from \eqref{effmk}, which gives
\be 
M_k=
\begin{pmatrix}
\gamma_k && \xi_k \\ 
-\xi_k && \gamma_k
\end{pmatrix},
\label{mknew}
\ee
where $\gamma_k$ is the effective friction given in \eqref{lambdanew} and
\be
\xi_k = \xi + \frac{k\epsilon^2}{2}
\label{rot}
\ee
is the effective rotation. As a result, we can immediately deduce from (\ref{ptrans}) and (\ref{Jtrans}) that
\be
p^*_k(\bx) = \frac{\gamma_k}{\pi\epsilon^2} \exp \left(-\frac{\gamma_k}{\epsilon^2}\|\bx\|^2 \right)
\label{eqefftransdens}
\ee
and 
\be 
\bJ_k^*(\bx) = \xi_k \begin{pmatrix}-y \\ x \end{pmatrix}p_k^*(\bx).
\label{effcurtransstoc}
\ee
Note that $\xi_k$ in \eqref{rot} is consistent with the rotational constraint
\be
\nabla\times (D^{-1}\bF_k) = \nabla\times (D^{-1}\bF) + k\nabla\times(\Gamma\bx),
\ee
found for the effective process (see Sec.~5.6 of \cite{chetrite2014}), which implies here a constraint on the skew difference of $M_k$:
\be
(M_k)_{1,2}-(M_k)_{2,1}= M_{1,2}-M_{2,1}+\epsilon^2k.
\ee

The stationary density and current obtained in \eqref{eqefftransdens} and \eqref{effcurtransstoc} are illustrated in Fig.~\ref{fig:fig4} for different fluctuation regimes associated with different $k$ values. The physical interpretation of these regimes follows closely our discussion of the gradient diffusion, keeping in mind that the transverse diffusion now has an inherent rotation that favors trajectories that rotate in the direction of the drift or stationary current. Indeed, taking the case $\xi > 0$, we observe the following:

1- Area fluctuations $a>a^*$ larger than the typical value, associated with $0<k<k_+$, are created by rare trajectories that rotate slightly faster than the natural rotation induced by the drift ($\xi_k>\xi$) and that are weakly attracted to the origin ($\gamma_k<\gamma$), so they venture further in the plane to accrue more area; see Fig.~\ref{fig:fig4}(a). As in the gradient case, the latter mechanism dominates for very large area fluctuations and is responsible for the exponential distribution seen for $a\gg a^*$. Moreover, $\xi_k$ also saturates to a maximum rotation when $k\ra k_+$ or, equivalently, when $a\ra\infty$, found from \eqref{range} and \eqref{rot} to be equal now to $\sqrt{\gamma^2+\xi^2}$ \footnote{We note that $\gamma_{k_a}$, where $k_a$ is the root of \eqref{legdiff}, is even as a function of $a$, while $\xi_{k_a}$ is odd in $a$.}.

2- Positive area fluctuations below $a^*$, associated with the range $-2\xi/\epsilon^2 < k < 0$, are created by trajectories that rotate less ($\xi_k < \xi$) and are closer to the origin ($\gamma_k > \gamma$), compared to typical trajectories of the system associated with $a^*$; see Fig.~\ref{fig:fig4}(b). The combination of the two effects leads to Gaussian small fluctuations close to $a^*$.

3- Negative area fluctuations, associated with $k_-<k < -2\xi/\epsilon^2$, are created by trajectories that rotate against the current ($\xi_k<0$), so there is a current reversal, though $\gamma_k$ is still reduced compared to $\gamma$, so those trajectories have larger excursions from the origin; see Fig.~\ref{fig:fig4}(c). As in the first fluctuation regime, the latter mechanism is dominant for very large negative fluctuations, explaining the exponential distribution seen for $a\ll 0$, since $\xi_k$ saturates to $-\sqrt{\gamma^2+\xi^2}$ when $k\ra k_-$, and so when $a\ra-\infty$.

Compared to the gradient diffusion, there is a further interesting effect in the transverse diffusion arising when we consider the value $a=0$ between the second and third fluctuation regimes above, associated with $k=-2\xi/\epsilon^2$. In this case, $\xi_k=0$ so the current of the effective process vanishes, which means that this process is reversible. This is expected physically, since the trajectories responsible for the rare event $A_T=0$ must effectively cease to rotate in a preferred direction in the long-time limit and, therefore, must appear in simulations or experiments as if they were produced by a reversible process. This is an important observation, showing that the stochastic area can be used as an irreversibility metric but only in a probabilistic way---from the measurement or estimation of $A_T$, one cannot say with certainty that a process is irreversible, but can only do so with a probability or confidence determined from the statistics of this quantity, as we discuss next. 

\section{Irreversibility test}

The mean stochastic area $A_T$ is a natural estimator of the asymptotic mean $a^*$, which can be used, as discussed, to determine whether a diffusion is reversible ($a^*=0$) or irreversible ($a^*\neq 0$) \cite{gonzalez2019}. As a statistical estimator, $A_T$ is asymptotically unbiased as well as consistent, since its variance decreases with the observation time $T$, so that $A_T$ converges in probability to $a^*$ as $T\ra\infty$. Naturally, for a finite time, the distribution of $A_T$ is not fully concentrated on $a^*$ -- it has a finite width around this value that determines the error bar or confidence interval associated with the estimation of $a^*$ and, in turn, the confidence that one has in deciding from observed trajectories that a diffusion is irreversible.

This confidence interval is obtained in the usual way from statistics using the fact that $A_T$ has Gaussian fluctuations around $a^*$ with the variance shown in \eqref{asympvar} \cite{asmussen2007}. Accordingly, if we set the confidence level at one standard deviation, then we can say that the area $A_T$ estimated from one trajectory of duration $T$ will lie within the interval $[a^*-\sigma_T,a^*+\sigma_T]$ with 67\% probability, where
\be
\sigma_T = \sqrt{\text{var}[A_T]} \sim \frac{\sqrt{\lambda''(0)}}{\sqrt{T}}
\ee
is the theoretical standard deviation of $A_T$. In practice, $a^*$ is estimated from $A_T$, while $\lambda''(0)$, the theoretical asymptotic variance, is estimated either from a single trajectory using batch mean methods \cite{bratley1987} or from a collection of many trajectories, copies or replicas, as commonly done in Monte Carlo simulations \cite{kalos2008}.
 
\begin{figure}[t]
\includegraphics{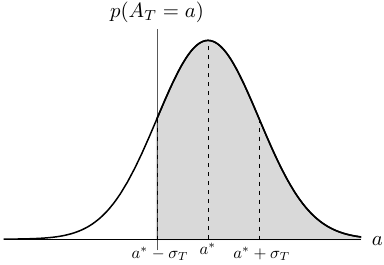}
\caption{Statistical test for irreversibility based on the probability that $A_T>0$ (in gray).}
\label{figprob}
\end{figure}

By extending this result to a one-sided confidence interval, illustrated in Fig.~\ref{figprob}, we can define a statistical test to determine whether a diffusion is irreversible \footnote{Note that irreversibility cannot be tested statistically by observing the event $A_T\neq 0$, since the probability of this event is trivially 1. Rather, we must use the fact that $a^*\neq 0$ induces an asymmetry in the fluctuations of $A_T$ to test for the event $A_T>0$ or $A_T<0$, depending on whether $a^*>0$ or $a^*<0$, respectively.}. To be specific, consider the case where $a^*>0$ and let $\sigma_T$ be such that $a^*-\sigma_T=0$, as in Fig.~\ref{figprob}. Then $P(A_T>0)\approx 0.84$ based on the Gaussian approximation. From the figure, it is also clear that, if $a^*-\sigma_T>0$, that is, if $0$ is outside  the one-sided interval, then $P(A_T>0)>0.84$ \footnote{This is a rough estimate, obviously, coming from the Gaussian approximation of $p(A_T=a)$, which can be refined knowing from our results that the tails of that density are exponential.}. The exact value of $P(A_T>0)$ in this case depends on $a^*$ and $\sigma_T$. A similar result applies if $a^*<0$ by reflecting the figure horizontally. In that case, we have $P(A_T<0)\approx 0.84$ if $a^*+\sigma_T=0$ and $P(A_T<0)>0.84$ if $a^*+\sigma_T<0$.
 
With these results, we propose the following steps for testing the irreversibility of a diffusion:

1- Simulate or observe a trajectory $(X_t)_{t=0}^T$ over the duration $T$ to estimate $a^*$ with $A_T$;

2- From the same trajectory, estimate the standard deviation $\sigma_T$ using batch means methods or, alternatively, re-generate many trajectories of duration $T$ to estimate that standard deviation;

3- Denote the estimator of $\sigma_T$ by $\hsigma_T$. If $A_T-\hsigma_T>0$ or $A_T+\hsigma_T<0$, then one is at least 84\% sure that the observed diffusion is irreversible;

4- If the test in Step 3 is inconclusive, then repeat the process for a longer duration $T$.

\begin{figure}
\includegraphics{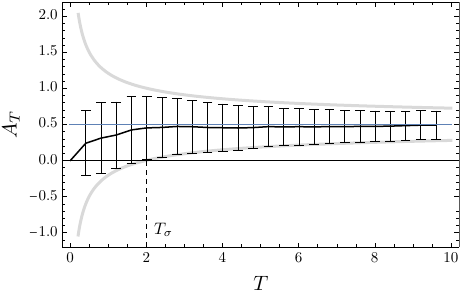}
\caption{Experimental test for determining whether a diffusion is irreversible based on the estimation of $a^*$. The estimated area $A_T$ (solid black line) is obtained from a trajectory of the transverse diffusion with $\gamma=\xi=\epsilon=1$, for which $a^*=0.5$ (blue line). As $A_T$ approaches $a^*$, the corresponding error bar gets above $0$ at some time $T_\sigma$ (dashed line), from which we can say that the system is irreversible with a probability greater than $0.84$. The gray line shows the theoretical standard deviation around the expectation $a^*$.}
\label{figtest}
\end{figure}

We illustrate this test in Fig.~\ref{figtest} for the transverse diffusion with $\gamma=\xi=\epsilon=1$ for which $a^*=0.5$, as seen from \eqref{asympmeantrans} and in Fig.~\ref{fig:fig3}. By simulating one trajectory of that system, we see that the estimated area $A_T$ fluctuates initially and starts to hover around $0.5$, so we start to be confident that the system is irreversible, based only on the observation of $A_T$. To quantify that confidence, we must compute the error bar associated with $A_T$ and wait until a later time, denoted by $T_\sigma$ in the plot, when the error bar stops containing the value $a=0$, thus rejecting the possibility that the system is reversible. From that point on in time, we are $84\%$ confident that the system is irreversible.
 
Note that, for simplicity, we have computed the error bars reported in Fig.~\ref{figtest} by simulating $N=100$ independent trajectories. In this case, the standard deviation of the estimated area is further reduced by a factor $\sqrt{N}$, since we have effectively collected a sample of 100 area values that are independent. In cases where this cannot be done, e.g., in an experimental setting where only one trajectory can be observed, the standard deviation of the estimated area can still be estimated using batch means methods, which proceed by splitting the trajectory in near-independent segments; see \cite{bratley1987} for details.

In the experimental test above, it is important to note that Step 3 being inconclusive does not mean that the system is reversible -- it only means that we cannot say at that point that the system is irreversible. In fact, it is not possible, strictly speaking, to decide that a system is reversible, even with some probability, because the latter case corresponds to the event $A_T=0$, which for a real random variable has zero probability to happen, and so cannot be tested statistically. 

This shows that, to test reversibility, one needs to access quantities or statistics other than $A_T$ that relate to the reversibility of a diffusion. One such quantity or indicator is the asymmetry between the fluctuating and relaxing parts of fluctuation loops observed in irreversible systems \cite{ghanta2017}. Another quantity that we propose for future studies is the skewness of $A_T$, which we expect to be generally zero for reversible diffusions and non-zero for irreversible diffusions, following the linear diffusions studied in the previous section. 
 
\section{Conclusion}

We have studied the statistics of the area enclosed by random trajectories of linear diffusions in two dimensions, following previous studies of this quantity initiated by L\'evy for planar Brownian motion. We have derived an exact form for the generating function of the stochastic area, and have derived from this function explicit expressions for the large deviation functions characterizing the probability density of the stochastic area in the long-time limit. The results show two different regimes of fluctuations around the stationary expected area: a \emph{Gaussian} regime of \emph{small} fluctuations near the expected area, which is important as we have seen for determining whether a diffusion is irreversible, and an \emph{exponential} regime of \emph{large} area fluctuations, which is symmetric or asymmetric when the underlying diffusion is reversible or irreversible, respectively. The difference between these two regimes reflects, as we have also seen, different types of rare trajectories. Small area fluctuations, on the one hand, tend to be created by trajectories that stay close to the origin but rotate at a different rate around this point, whereas large area fluctuations, on the other, tend to be created by trajectories that venture far from the origin as a way to accrue more area while rotating around this point at a more or less constant rate.

Similar results hold for other linear diffusions, including the Brownian gyrator \cite{monthus2022,buisson2022,buisson2022b}, which is driven in a nonequilibrium state by a temperature difference \cite{filliger2007,dotsenko2013,argun2017,chiang2017,viot2023}, as well as systems involving correlated noise components \cite{kwon2011}, which lead to non-diagonal components in $D$. The methods that we have presented can also be applied to higher-dimensional linear diffusions by considering the stochastic area $(A_T)_{ij}$ in all possible planes $(i,j)$ of the $\reals^n$ space, so as to define a stochastic area matrix or \emph{tensor} \cite{ghanta2017}. This tensor is antisymmetric by definition, and so has $n(n-1)/2$ independent components, which means that testing the irreversibility of a diffusion in $\reals^n$ requires, in principle, that we apply the test proposed here for that many components. This can be a difficult task, but it is simpler, nevertheless, than estimating the current field $\bJ^*(\bx)$ over a grid in $\reals^n$ from observed trajectories. 

Finally, we expect many of the properties of the stochastic area found here for linear models to hold more generally for nonlinear ergodic diffusions. In particular, the distribution of the area should be asymmetric whenever the underlying diffusion is irreversible, leading to a skewness of the area, also observed recently for other current-type observables and processes \cite{wampler2022,salazar2022,manikandan2022}. Similarly, the two regimes, Gaussian and exponential, of fluctuations that we have found should apply more generally to nonlinear diffusions and should come again from the competition that exists between rotating and wandering trajectories. In this context, it would be interesting to see if there is a general bound on the rotation of trajectories associated with large area fluctuations, as found here for gradient and transverse diffusions, which would imply a bound on their vorticity.

\begin{acknowledgments}
T.D.P.M. was funded by the Science Faculty of Stellenbosch University (MSc Scholarship). J.d.B. was funded by South Africa's NRF (PhD Scholarship). 
\end{acknowledgments}

\bibliography{masterbib}

\end{document}